\documentclass[preprint,showpacs,superscriptaddress,preprintnumbers,amsmath,amssymb]{revtex4-1}
\usepackage{graphicx}
\usepackage{bm}
\usepackage{amsmath}
\usepackage{dcolumn}
\usepackage{dsfont}
\usepackage{amssymb}
\usepackage{tabularx}
\usepackage{array}
\usepackage{float}
\usepackage{placeins}
\usepackage{color}
\usepackage{epstopdf}
\usepackage{mathrsfs}

\usepackage[colorlinks, linkcolor=blue,anchorcolor=blue,citecolor=blue,urlcolor=blue]{hyperref}

\begin{document}
\draft
\title {Emergence of an upper bound to the electric field controlled  Rashba spin splitting in InAs nanowires}
\author {Jun-Wei Luo}
\email{jwluo@semi.ac.cn}
\affiliation{State key laboratory of superlattices and microstructures, Institute of Semiconductors, Chinese Academy of Sciences, Beijing 100083, China}
\affiliation{Center of Materials Science and Optoelectronics Engineering, University of Chinese Academy of Sciences, Beijing 100049, China}
\affiliation{Beijing Academy of Quantum Information Sciences, Beijing 100193, China}
\author {Shu-Shen Li}
\affiliation{State key laboratory of superlattices and microstructures, Institute of Semiconductors, Chinese Academy of Sciences, Beijing 100083, China}
\affiliation{Center of Materials Science and Optoelectronics Engineering, University of Chinese Academy of Sciences, Beijing 100049, China}
\author{Alex Zunger}
\address{Energy Institute, University of Colorado, Boulder, Colorado 80309, USA}

\date{\today}

\begin{abstract}
The experimental assessment of the strength ($\alpha_R$) of the Rashba spin-orbit coupling is rather indirect and involves the measurement of the spin relaxation length from magnetotransport, together with a model of weak antilocalization. The analysis of the spin relaxation length  in nanowires, however, clouds the experimental assessment of the $\alpha_R$ and  leads to the prevailing belief that it can be tuned freely with electric field--a central tenant of spintronics. Here, we  report direct theory of $\alpha_R$ leading to atomistic calculations of  the spin band structure of InAs nanowires upon application of electric field-- a direct method that does not require a theory of spin relaxation. Surprisingly, we find an  {\it upper bound} to the electric field tunable Rashba spin splitting and the ensuing  $\alpha_R$; for InAs nanowires,  $\alpha_R$ is  pinned at about 170 meV{\AA} irrespective of the applied field strength. We find that this pinning is due to the quantum confined stark effect, that reduces continuously the nanowire band gap with applied electric field, leading eventually to band gap closure and a considerable increase in the density of free carriers. This  results in turn in a strong screening that prevents the applied electric field inside the nanowire from increasing further beyond around 200 kV/cm for InAs nanowires.  Therefore, further increase in the gate voltage will not increase $\alpha_R$. This finding clarifies the physical trends to be expected in nanowire Rashba SOC and the roles played by the nano size and electric field.
\end{abstract}

\maketitle

Spintronics offers the use of electron spin rather than electron charge to carry information, whereby the needed magnetic field is effectively provided by Rashba spin-orbit coupling (SOC)  \cite{dresselhaus55, bychkov84, zhang14} rather than by external magnetic field  \cite{nowack07,manchon15}. This opens a  route towards electrical manipulation of electron spins  \cite{awschalom09}, such as that proposed  in the Datta-Das spin transistor \cite{datta90} and spin qubits \cite{nadj-perge10,nowack07}.  Such electrical manipulation  instead of magnetic manipulation is particularly appealing for  this purpose, because electric fields are easy to create locally on-chip, simply by exciting a local gate electrode \cite{nowack07}.   


One-dimensional semiconductor nanowires with strong SOC have recently emerged as  promising building block for spintronics \cite{nadj-perge10} and as  a unique solid state platform for realizing and observing the  Majorana fermions \cite{oreg10, lutchyn10, mourik12, das12}. However,  despite such interest of SOC in 1D wires, they have been studied far less than in 3D bulk semiconductors \cite{dresselhaus55, luo09, bulkRashba11, zhang14} and  in 2D heterostructures  \cite{bychkov84, winkler03} and quantum wells \cite{winkler03,luo10}. Here we focus on the assessment of the strength of the Rashba SOC ($\alpha_R$) and its dependence on the applied external electric field. The prevailing practice to deduce Rashba parameter $\alpha_R$ is to measure the spin relaxation length $l_{\text{so}}$ from magnetotransport measurements requiring an  analysis of weak antilocalization  \cite{dhara09, estevez10,liang12, van15}.  Considering that the D'yakonov-Perel' (DP) mechanism a primary  spin-relaxation in 3D bulk and 2D quantum wells is considerably suppressed in 1D nanowires \cite{kiselev_suppress_2000,pareek_suppress_2002, pramanik_suppress_2005, kaneko_suppress_2008, holleitner_suppress_2006, kettemann_suppress_2007} and the inter-subband scattering induced spin-relaxation becomes dominate,  the experimentally deduced Rashba parameter $\alpha_R$ in 1D nanowires  from magnetotransport measurements  may be uncertain because unlike the former the latter is independent on $\alpha_R$.

In this work, we provide a theoretical evaluation of $\alpha_R$ in InAs nanowires using a direct method that does not require a theory of spin relaxation.  We do so by solving the Schr\"odinger equation for a nanowire experiencing a perpendicular electric field, obtaining the spin-dependent band structure (see Fig. 1(b) for a 30 nm InAs nanowire), from which we directly obtain the spin splitting by subtracting the band energies of the branches with two spin directions.  We then fit the obtained spin splitting  $\Delta E_{ss}(k)$ of the lowest conduction subband to a wavevector power series: $\Delta E_{ss}({ k})=2\alpha_R k+\gamma_R k^3$ and thus  find directly the Rashba parameter $\alpha_R$. We study electron $\alpha_R$ for InAs  nanowires as a function of nanowire size and electric field.  Our central finding is that there is an upper bound  to the strength $\alpha_R$  of the field-induced Rashba SOC owing to the quantum-confined stark effect (QCSE). This finding explains the size-independence and field-independence of $\alpha_R$ = 200 meV{\AA} of InAs nanowires observed in a recent experiment \cite{roulleau10}.  This finding clarifies the physical trends to be expected in nanowire Rashba SOC, sets realistic expectations to nanowire spintronics applications, and resolves the experimental puzzle of occasional failure to raise persistently $\alpha_R$ with gate voltage  in  nanowires \cite{hansen05, dhara09, estevez10, roulleau10, liang12}.

\paragraph*{Atomistic calculation of the Rashba Spin splitting in nanowires under applied electric field.}
In the ${\bf k\cdot p}$ approaches for studying the Dresselhaus and Rashba SOC in low-dimensional structures \cite{kloeffel11, zhang06, zhang07, winkler03},  one uses a phenomenological Hamiltonian where one needs to decide at the outset which 3D bulk bands couple in low-dimensional structures by the SOC and crystal field. The potential of missing important physical interactions, such as heavy-hole and light-hole band coupling, not selected to be present in certain model Hamiltonian can be substantial to describe SOC induced spin splitting \cite{luo11, luo10,luo10b}.  Here we adopt instead an atomistic pseudopotential method in which the low-dimensional structure is viewed as a giant molecular system in its own right, rather than express it in terms of a pre-selected basis drawn from a reference 3D bulk system.  
This method has been tested extensively over the past two decades for a broad range of spectroscopic quantities in self-assembled and colloidal nanostructures \cite{bester05c, luo09b,luo15}, as well as been previously applied to investigate Dresselhaus SOC  in 3D zinc-blends semiconductors \cite{luo09}, 2D  quantum wells \cite{luo10, luo10b}, and 1D nanowires \cite{luo11}.

The band structure of 1D nanowire is obtained via direct-diagonalization of the  Schr\"{o}dinger equation  \cite{luo11,luo09,bester05c},
\begin{equation}
\Big(-{1\over 2}\nabla^2+V({\bf r})+|e|{\bf E}\cdot {\bf r}\Big)\psi_i({\bf r})=\epsilon_i\psi_i({\bf r}).
\end{equation}
The crystal potential $V({\bf r})=\sum_{n,\alpha}\hat{v}_{\alpha}({\bf r}-{\bf R}_{n,\alpha})$
is a superposition of screened atomic potentials $\hat{v}_{\alpha}$ of atom type $\alpha$ located
at atomic site ${\bf R}_{n,\alpha}$. The screened atomic potential $\hat{v}_{\alpha}$  contains a local part
$v^{L}_{\alpha}$ and a nonlocal spin-orbit interaction part
$\hat{v}^{NL}_{\alpha}$ which is treated as local in the Kleinman-Bylander scheme \cite{kleinman82}. ${\bf E}=(E_x, E_y, E_z)$ is applied electric field \cite{bester05c, yu05, winkler03}, which is generally created in devices by exciting a local gate electrode \cite{nowack07}. Here,  ${\bf E}=(E_x, 0, 0)$ applied in the $x$-direction, perpendicular to the nanowire axis $z$-direction. The construction of the screened pseudopotenttial $\hat{v}_{\alpha}$ is the key to accuracy and realism. To remove the ``LDA error'' in the bulk crystal we fit the atomic potentials $\hat{v}_{\alpha}$ to experimental transition energies, effective masses, spin-orbit splitting, and deformation potentials of the parent bulk semiconductors as described previously \cite{InAs_EPM_2000}.  The InAs nanowires are embedded in an artificial material with the same lattice as InAs but much wider bandgap and heavier masses \cite{wang_reinterpretation_2015}. For more details on screened pseudopotentials of InAs and barrier material used here see Refs. \cite{InAs_EPM_2000, wang_reinterpretation_2015}.

\paragraph*{Closed form physical model for Rashba electron $\alpha_R$ term in nanowires.}
A comparison of $\alpha_R$ between atomistic pseudopotential calculations and the classical model Hamiltonian approach may provide insight into the understanding of  the  Rashba  spin splitting.  The term of the Rashba SOC, which originates from the spin-orbit interaction,  in an effective $2\times 2$ $\Gamma_{6c}$ conduction band Hamiltonian is arising from the non-commutativity of wavevector $k$ and crystal potential $V$ from a decoupling of conduction and valence band states \cite{winkler03}.  If one uses the $8\times 8$ Kane Hamilotonian, third-order perturbation theory for the conduction band Hamiltonian yields the Rashba SOC term  \cite{winkler03},
\begin{equation}
H_R=r_{41} \boldsymbol {\sigma}\cdot {\bf k}\times {\bf E},
\end{equation}
where ${\bf E}=(1/e)\nabla V$ is the electric field contained implicitly in the crystal potential $V$, here, ${\bf E} = (E_x,0,0)$ is arising from applied electric field $E_x$. $\boldsymbol {\sigma}=(\sigma_x, \sigma_y,\sigma_z)$  the vector of Pauli spin matrices, and a material-specific  Rashba coefficient \cite{winkler03},
\begin{equation}
r_{41}=\frac{aeP^2}{3}\Big[\frac{1}{{E_g}^2}-\frac{1}{(E_g+\Delta_{\text{so}})^2}\Big].
\end{equation}
Where $e$ is electron charge, $E_g$ the band gap of quantum structures, $\Delta_{\text{so}}$ the spin-orbit splitting ($\Delta_{\text{so}}=0.38$ eV for bulk InAs), and $P$  Kane's momentum matrix element ($P^2/2m=21.5$ eV for bulk InAs). The adjustable parameter $a$ is used to take into account all factors missed in classical Hamiltonian approach, such as (i) inter-band coupling induced by space confinement rather than  by ${\bf k\cdot p}$; (ii) Change of the quantum confinement potential  induced modification of dipole matrix element $P$;  (iii) Energy level splitting of the valence bands; and (iv) QCSE as discussed below. The Rashba SOC induced spin splitting  of the conduction subband is $\Delta \epsilon(k)=2\alpha_R k$, here the pre-factor is the named Rashba parameter,
\begin{equation}
\label{1Drashba}
\alpha_R=r_{41}E_x=\frac{aeP^2E_x}{3}\Big[\frac{1}{{E_g}^2}-\frac{1}{(E_g+\Delta_{\text{so}})^2}\Big].
\end{equation}

The value of $a$ could be inferred by fitting atomistic predicted $\alpha_R$ to  Eq. (\ref{1Drashba}). In doing so, we first describe atomistic predicted $E_g(D)$ of InAs nanowires with diameter $D$ by a formula $E_g(D)=E^b_g+\beta /D^{\gamma}$ as shown in Fig. 3 (a), here $E^b_g=0.417$ eV is bulk  InAs band gap \cite{vurgaftman01}, and $\beta=8.49$ and $\gamma=1.58$ for InAs nanowires under $E_x=30$ kV/cm.  Taking $a$  as the only adjustable parameter, we subsequently fit  atomistic predicted $\alpha_R(D)$ to Eq. (\ref{1Drashba}), shown in Fig. 2 (a). Our best fit indicates $a=0.92$. This value is close to $a=0.96$ for the rectangular well  and $a=1$  for the parabolic well  \cite{winkler03}, with an infinite energy barrier \cite{winkler03},  implying that the contributions from missed four factors to $\alpha_R$ are small, at least for nanowires under a moderate electric field of $E_x=30$ kV/cm.

\begin{figure}[!hbp]
\includegraphics[width=0.8\textwidth]{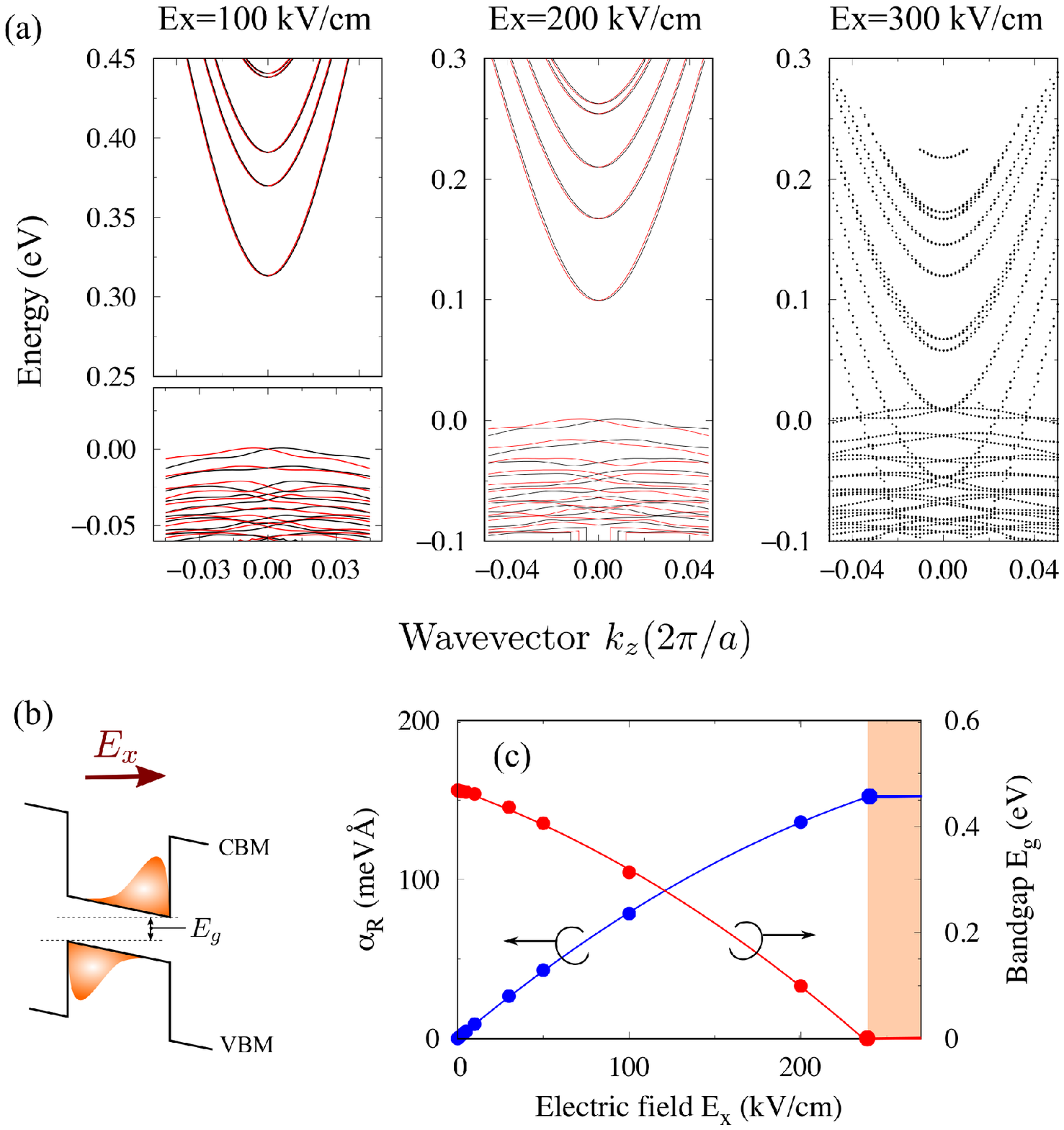}
\caption{\label{fig:QCSE}  (a) Schematic  illustration of a nanowire under a perpendicular electric field. (b) Band structures of the D=30 nm InAs nanowire under three electric fields of 100, 200, and 300 kV/cm, respectively. (b) Band diagram and wave function distribution of the nanowire under a perpendicular electric field. (c) Bandgap and Rashba parameter $\alpha_R$ of the D=30 nm InAs nanowire  as increasing the electric field.}
\end{figure}

\paragraph*{Dependence of  $\alpha_R$ on wire diameter and applied field:  the emergence of saturation.}
Our recent  work \cite{luo11} on spin splitting in zinc-blende nanowires established based on fundamental nanowire symmetry that  the Dresselhaus spin splitting is absent in the (001)- and (111)-oriented nanowires with such tetrahedral bonding. An electric field applied perpendicular to the wire direction breaks the symmetry but does not evoke the Dresselhaus spin splitting, even if such spin splitting is present in 3D bulk InAs and 2D quantum wells. The field-induced spin splitting is exclusively due to the Rashba SOC effect \cite{luo11}.  In the following we apply  the electric field perpendicular to the (001)-oriented InAs zinc-blende nanowires; the obtained spin splitting (Fig.~\ref{fig:QCSE} (a)) is thus exclusively due to the Rashba effect. In early reports \cite{glas_why_2007, johansson_diameter_2010}, small ($<100$ nm) diameter III-V nanowires showed a tendency toward forming a wurtzite phase. In recent reports, however, both pure ZB and pure WZ nanowires could be achieved across the broad range of nanowire diameters \cite{joyce_phase_2010, zhao_growth_2014, fu_crystal_2016}. Zinc-blende InAs nanowires as small as 15 nm in diameter were routinely synthesized \cite{fu_crystal_2016}.

\begin{figure}[!hbp]
\includegraphics[width=0.8\textwidth]{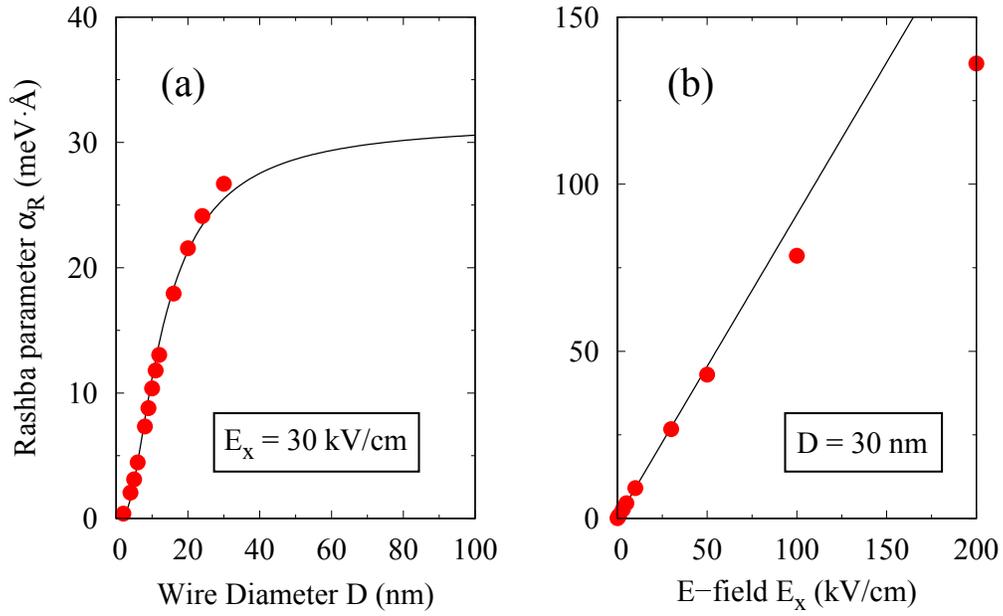}
\caption{\label{fig:Rashba}  The strength of the Rashba effect ($\alpha_R$) in InAs nanowires obtained from atomistic pseudopotential method calculations. (a) Under a fixed electric field $E_x=30$ kV/cm, Rashba parameter $\alpha_R$ as a function of nanowire diameter $D$ predicted by the atomistic pseudopotential approach  (filled circles). The best fit of atomistic predicted $\alpha_R$ to Eq. (3)   is indicated by a solid line with $\beta=8.49$ and $\gamma=1.58$.  (b) Rashba parameter $\alpha_R$ as a function of electric field for a $D=30$ nm nanowire. }
\end{figure}

Fig.~\ref{fig:Rashba} shows atomistic calculated  $\alpha_R$ in InAs nanowires as a function of wire diameter $D$ for a fixed electric field. Upon application of a fixed electric field $E_x=30$ kV/cm,  Fig.~\ref{fig:Rashba} (a)  exhibits that $\alpha_R$  increases rapidly to 21.5 meV{\AA}  up to $D=20$ nm and begins to saturate to 34 meV{\AA} (a value of bulk InAs) as further increasing the nanowire diameter for InAs nanowires.  The best fit of atomistic predicted $\alpha_R$ to Eq. (4), as shown in Fig. 2 (a) by a back curve, indicates a good agreement between atomistic method and classical model Hamiltonian approach for Rashba SOC. Fig.~\ref{fig:Rashba} (b) shows the field-dependence of $\alpha_R$ for a $D=30$ nm InAs nanowire which is linear  until $E_x=50$ kV/cm and then becomes sublinear as  further increasing $E_x$. It clearly manifests that the Rashba SOC is  strongly field tunable:  $\alpha_R$ increases from zero at the absence of electric field to as large as 136 meV{\AA} at $E_x=200$ kV/cm. The slope in the linear region determines the Rashba coefficient $r_{41}=\partial \alpha_R/\partial E_x=90.9$ e\AA$^2$, which is consistent with bulk InAs of  $r_{41}=117.1$ e\AA$^2$ \cite{winkler03} with a small difference owing to quantum confinement effect, indicating the robustness of the used atomistic pseudopotential method to predict the Rashba effect.

\begin{figure}[!hbp]
\includegraphics[width=0.8\textwidth]{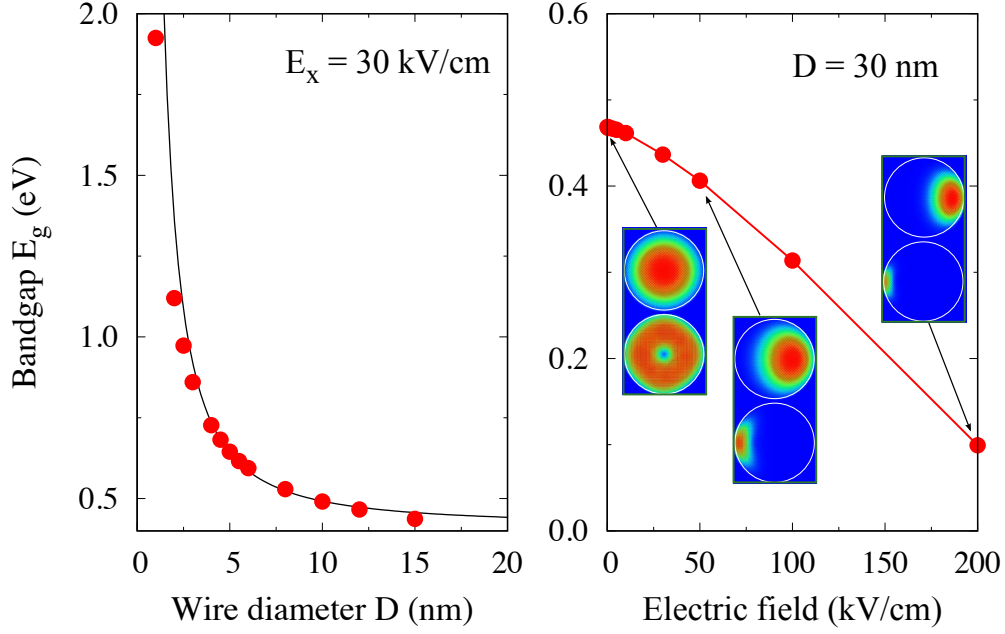}
\caption{\label{fig:InAsEg}  (a) Atomistic method predicted bandgap $E_g$ as a function of nanowire diameter $D$ for InAs nanowires upon application of an electric field $E_x=30$ kV/cm. The solid line represents the best fit of the nanowire $E_g$.  Inset to (a) ratio of $\alpha_R$ predicted by the atomistic method and classical approach, respectively. (b) Nanowire bandgap $E_g$ as a function of electric field for the $D = 30$ nm InAs nanowire. Insets show the wave function square of CBM and VBM for a $D = 30$ nm InAs nanowire responding to $E_x=0$,  $E_x=50$, and  $E_x=200$ kV/cm, respectively.}
\end{figure}

\paragraph*{Effect of electric field on $\alpha_R$  through electron-hole charge separation: the QCSE.}
When an external electric field is applied perpendicularly to a nanowire, the electron states shift to lower energies, while the hole states shift to higher energies,  reducing the nanowire band gap $E_g$, as shown in Fig.~\ref{fig:InAsEg} (b). Additionally, the external electric field shifts electrons and holes to opposite sides along the electric field within the nanowire cross-section, see insets to Fig.~\ref{fig:InAsEg} (b) and Fig.~\ref{fig:QCSE} (b), decreasing the overlap integral, which in turn reduces the recombination efficiency (i.e., fluorescence quantum yield) of the system. This effect is the so-called QCSE.  The QCSE modifies explicitly   $\alpha_R$ via shifting the energy levels, at the same time, changes implicitly  $\alpha_R$ by reducing the overlap of the wave functions of the conduction and valence subbands and subsequently decreasing the matrix elements. We expect stronger QCSE under the larger electric field.  To examine the modification of $\alpha_R$  induced by QCSE  in nanowires, we investigate  the evolution of the $\alpha_R$ as a function of electric field for a $D=30$ nm nanowire, as shown in Fig.~\ref{fig:Rashba} (b). It exhibits the field-dependence of the $\alpha_R$  being sublinear instead of expected linear  from the classical model Hamiltonian approach (will be discussed below).  Specifically, the field-dependence of the $\alpha_R$ is almost linear  until $E_x=50$ kV/cm and then becomes sublinear as further increasing $E_x$. The sublinear behavior is a result of QCSE, and a larger deviation from the linear function of  $\alpha_R(E_x)$  illustrates a stronger QCSE on $\alpha_R$. This result is consistent with  what we have discussed above that the QCSE  on $\alpha_R$ is negligible under $E_x=30$ kV/cm.

\paragraph*{Emergence of an upper bound for Rashba parameter  $\alpha_R$ in nanowires.}
Figure~\ref{fig:InAsEg} (b) shows the QCSE induced shifting of the band gap $E_g$ for a $D=30$ nm InAs nanowire. We see that the QCSE shifts  the nanowire $E_g$ to a smaller value continuously and finally to as low as 0.05 eV at $E_x=200$ kV/cm. Further increase of $E_x$ will ultimately close the bandgap and make the nanowire metallic, which leads to considerable increase in free carrier density in the nanowires and produces a giant screening which in turn prevents the electric field inside the nanowire from further increasing. In experimental devices, the magnitude of the electric field applied across the nanowires are tuned indirectly  by a gate voltage. Although one  may increase the gate voltage  as large as to tens Volts \cite{liang12},  the  electric field falling inside the nanowire is pinned to a value once it closes the nanowire bandgap owing to the QCSE. To further increase gate voltage above pinned electric field, the additional voltage  will drop across the matrix outside the nanowires. Therefore, the QCSE gives rise to an upper limit of the reachable electric field, which is around 200 kV/cm for InAs nanowires.  The predicted  $\alpha_R=136$ meV{\AA} at $E_x=200$ kV/cm is thus a maximum achievable value for the $D=30$ nm InAs nanowire, as shown in Fig.~\ref{fig:QCSE} (c).  Considering that $\alpha_R$ increases slightly  in thicker nanowires, we estimate an upper bound for $\alpha_R$ being  about 170 meV{\AA} for  InAs nanowires.

\paragraph*{Comparison between theoretical predictions and experimental measurements of Rashba parameter $\alpha_R$.}
The maximum available electric field and the strength of the Rashba SOC in nanowires have frequently observed in the experimental measurements  \cite{hansen05, dhara09, estevez10, roulleau10, liang12}, but has not recognized as essential physical effects. Regarding the classical model Hamiltonian,  Rashba parameter $\alpha_R$ is expected to be simply proportional to the magnitude of applied electric field. We thus believe that we could always increase the strength of the Rashba SOC via  increasing the gate voltage applied to the nanowires  \cite{liang12}. However, experiments often failed to realize it  \cite{hansen05, dhara09, estevez10, roulleau10, liang12}. We demonstrate from atomistic calculations that there exists an upper bound to $\alpha_R$ for each nanowires, thus clarifying the experimental puzzle of failure to raise persistently $\alpha_R$ with gate voltage   in  nanowires \cite{hansen05, dhara09, estevez10, roulleau10, liang12}.

The general practice to deduce the strength of  Rashba parameter $\alpha_R$ is to measure the spin relaxation length $l_{\text{so}}$ from magnetotransport measurements requiring an  analysis of weak antilocalization  \cite{dhara09, estevez10,liang12, van15}. In the dirty metal regime (where the electron elastic-scattering length $l_e$ is smaller than the wire diameter $D$) \cite{roulleau10}, the ballistic spin-precession length $l_R^2=l_{\text{so}}D/\sqrt{3}$, and subsequently $\alpha_R=\hbar^2/(2m^*l_R)$.  Experimentally reported $\alpha_R$ for InAs nanowires has covered  a wide range of $50-320$ meV{\AA}   \cite{hansen05, dhara09, estevez10, roulleau10, liang12}. Note that the wire geometries $D$ and expressions for $\alpha_R$ used by different groups vary  and that often only $l_{\text{so}}$  \cite{liang12, hansen05,dhara09,estevez10}, not $l_R$ \cite{roulleau10, van15}, is evaluated. Using $l_R$ to calculate $\alpha_R$ via $\alpha_R=\hbar^2/(2m^*l_R)$, Roulleau et al.  \cite{roulleau10} got a same Rashba coupling parameter of $\alpha_R$ = 200 meV{\AA}, under large gate voltage, for all three investigated InAs nanowires with diameters of 75,  140, and 217 nm, respectively. Such observed size-independence and field-independence of $\alpha_R$ = 200 meV{\AA}  supports well our prediction of the emergence of  an  upper bound of  $\alpha_R= 170$ meV{\AA} for InAs nanowires upon application of electric field.

We also note that a considerable suppression of the DP spin-relaxation, which usually dominates the spin-relaxation in semiconductors, in 1D nanowires has been predicted theoretically \cite{kiselev_suppress_2000,pareek_suppress_2002, pramanik_suppress_2005, kaneko_suppress_2008} and observed experimentally \cite{holleitner_suppress_2006, kettemann_suppress_2007}. The DP mechanism is owing to the randomizing of the momentum-dependent Rashba SOC-induced effective magnetic field.  In single-channel (i.e., single-band) clean 1D nanowires, where the electron elastic-scattering length $l_e$ is larger than the wire diameter $D$, the  spin-relaxation is even completely suppressed due to a dimensionally constrained DP mechanism  \cite{kiselev_suppress_2000, pramanik_suppress_2005,kaneko_suppress_2008}. Whereas, in multiple-channel 1D nanowires, the inter-subband scattering enables the spin-relaxation \cite{pramanik_suppress_2005, kaneko_suppress_2008}. Considering the inter-subband scattering induced spin-relaxation dependents mainly on the occupation of excited subbands, rather than on the strength of Rashba SOC, the experimentally deduced Rashba parameter $\alpha_R$ in 1D nanowires from magnetotransport measurements  may uncertain.

Interestingly, the conclusions regarding the Rashba parameter drawn from zinc-blende nanowires are also applicable to the wurtzite phase (as well as to different nanowire orientations and shapes). The reason for this is that the strength of the Rashba SOC for a given material dependents primarily on the nanowire band gap and is rather insensitive to geometric and crystal parameters that result in a particular band gap value. The excellent agreement between the atomistic method and classical model Hamiltonian on the prediction of $\alpha_R$, as shown in Fig. 2 (a), evidences it, since the model deduced $\alpha_R$ in Eq. (S3) applies to both 2D and 1D electrons \cite{winkler03, zhang06} without considering the crystal orientations and wire shapes.  Eq. (1) is well justified only for dielectric phase. Under the conditions when electron and hole densities develop at opposite edges of the wire, they produce additional electric potential, and, hence, a self-consistent Poisson-Schrodinger problem should in principle be solved. In other words, Eq. (1) is valid only until the dielectric stays constant, and narrowing of the gap seems to violate this condition. We did not perform such corrections. Indeed, we study charge neutral nanowires by specifying the Fermi level located in the middle of the band gap, and the applied electric field being uniform across the nanowire. A self-consistent simulation would depend on a range of device parameters such as the distance from the gate to the nanowire; the nature of the gate contact to the nanowire, the nature of the insulating layers between them, the doping levels and the dopant concentration inside the nanowire. Such quantities often vary from sample to sample and are not always cited in the experimental papers. Note, however, that the Rashba SOC is not dependent on the detailed profile of the local electric field because it is proportional to the expectation value of the external field $\alpha_R=<r_{41}E_x>$ as pointed out in Ref. \cite{winkler03}. Hence, we can approximate the Rashba SOC by solving only the Schrodinger equation with an assumption of a constant  (non-self-consistent) electric field across the nanowire cross-section. This approximation has been shown previously to be successful in reproducing the experimentally measured Rashba parameters as reported in literature \cite{winkler03}.

\paragraph*{Electron Rashba spin-orbit energy $E_{so}$  for Majorana detection experiments: } 
Semiconductor nanowires  were recently recognized to be a unique solid state platform for realizing and observing the  Majorana fermions -- unique particles that are identical to their own antiparticles, and forming bound states with non-Abelian exchange statistics and suitable as the building blocks of quantum computer \cite{stern13}.    Specifically, the Majorana fermions were recently detected within  the topological band gap of hybrid superconductor-semiconductor InAs  \cite{das12} or InSb nanowires  \cite{mourik12}, respectively, following the theoretical proposal \cite{oreg10, lutchyn10}. This experiment requires large SOC-induced Rashba spin splitting  since it determines the size of the topological gap that  needs to exceed $k_BT$ at the temperature at which experiment is performed.  Therefore, InAs or InSb nanowires were employed in Majorana fermion experiments considering bulk InAs and InSb have strong spin-orbit interactions  \cite{mourik12, das12}. 

\begin{figure}[!hbp]
\includegraphics[width=0.8\textwidth]{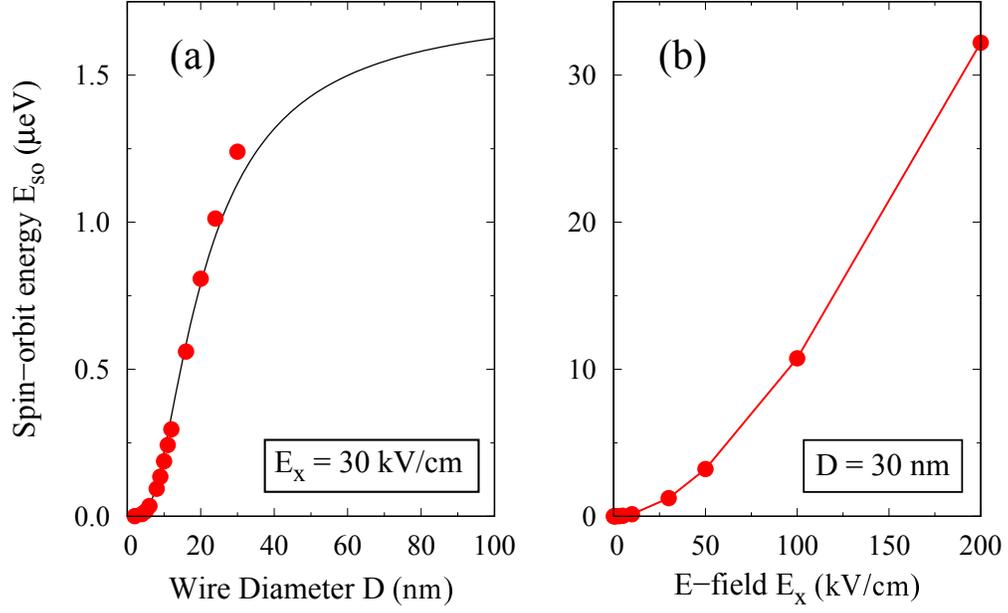}
\caption{\label{fig:InAsEso}  Spin-orbit energy $E_{so}$ of the lowest energy conduction subband in InAs nanowires. (a) Evolution of $E_{so}$ as a function of nanowire diameter $D$ for InAs nanowires  upon application of electric field $E_x=30$ kV/cm.  (b) Evolution of  $E_{so}$ as a function of electric field $E_x$ for a $D=30$ nm nanowire.}
\end{figure}

In Majorana fermion experiments, Rashba spin-orbit energy $E_{so}=m^*\alpha^2_R/2\hbar^2$  of the investigated nanowires is a critical parameter because it limits  the size of the applied magnetic field opened topological gap  that  needs to exceed the temperature at which the experiments  of Majorana fermions are carried out \cite{mourik12, das12}. $E_{so}$ is also estimated indirectly from experimental measurement. Here, we could directly obtain $E_{so}$ and its dependence on field and diameter from the calculations of $\alpha_R$ which is shown in Fig. 2 in the main text.  Fig.~\ref{fig:InAsEso} shows  $E_{so}$ of the lowest electron  subband for InAs nanowires.  We see that, under a moderate electric field of 30 kV/cm,   $E_{so}$ is unexpected  small (less than 2 $\mu$eV) and is far less than the minimum value required for hosting Majorana fermions.  A large electric field is thus expected to considerably increase  $E_{so}$ so as to meet the requirements of Majorana fermion experiments, such as Ref. \cite{das12} and Ref. \cite{mourik12} cite $E_{so}=70$  and 50 $\mu$eV for InAs and InSb nanowires, respectively. Instead, Fig.~\ref{fig:InAsEso}(c) shows that an extremely large electric field of $E_x=200$ kV/cm is needed to tune   $E_{so}$  to the maximal achievable value of 32 $\mu$eV  for a $D=30$ nm InAs nanowire.

In summary, we have studied the Rashba SOC directly in InAs nanowires by performing SOC band structure calculations using all-band atomistic pseudopotential approach, without unambiguous as occured  in experimental measurements. We uncovered the existence of  an upper bound of the strength of the electric field tunable Rashba SOC in semiconductor nanowires as increasing the gate voltage. We found that it is a result of the QCSE, which lowers the nanowire band gap as the applied electric field increases continuously, and finally, closes the band gap rendering the nanowire metallic. The metallic nanowires have a giant screening to prevent the electric field from further rising in the nanowire, and thus further increase the gate voltage will not increase the electric field, and thus $\alpha_R$.  The revealed upper bound of $\alpha_R=170$  meV{\AA} explains the size-independence and field-independence of $\alpha_R$ = 200 meV{\AA} of InAs nanowires observed in a recent experiment \cite{roulleau10}.  We believe that we have clarified  the experimental puzzle of failure to raise persistently $\alpha_R$ with gate voltage  in  nanowires \cite{hansen05, dhara09, estevez10, roulleau10, liang12}, making a fundamental step towards the understanding of the Rashba SOC in semiconductor nanowires.


\begin{acknowledgments}
The work in China was supported by the National Science Foundation of China (NSFC grants \#61888102 and \#11925407). AZ was supported by Office of Science, Basic Energy Science, MSE division under grant DE-FG02-13ER46959 to CU Boulder.
\end{acknowledgments}

\bibliography{nanowireRashba}

\FloatBarrier
\newpage

\FloatBarrier
\newpage

\end{document}